\documentstyle[prl,aps,preprint,floats]{revtex}

\begin{document}

\title{New solutions in 3D gravity}

\author{Yuri N.~Obukhov\footnote{On leave from: Dept. of Theoret. 
Phys., Moscow State University, 117234 Moscow, Russia}}
\address{Institute for Theoretical Physics, University of Cologne,
50923 K\"oln, Germany}

\maketitle

\begin{abstract}
We study gravitational theory in $1+2$ spacetime dimensions which is
determined by the Lagrangian constructed as a sum of the Einstein-Hilbert
term plus the two (translational and rotational) gravitational Chern-Simons 
terms. When the corresponding coupling constants vanish, we are left with 
the purely Einstein theory of gravity. We obtain new exact solutions for the 
gravitational field equations with the nontrivial material sources. Special 
attention is paid to plane-fronted gravitational waves (in case of the Maxwell 
field source) and to the circularly symmetric as well as the anisotropic 
cosmological solutions which arise for the ideal fluid matter source.
\end{abstract}
\bigskip


\section{Introduction}

Gravity in a three-dimensional spacetime attracts a lot of attention recently, 
see \cite{miebaek,tres92,kawai,cs,mil,aa} and the references therein. 
In a quite general setting, the gravitational field is described in terms
of the coframe 1-form $\vartheta^\alpha$ and the linear connection 1-form
$\Gamma_\alpha{}^\beta$. These ``potentials" give rise to the field strengths
which are the 2-forms of torsion $T^\alpha$ and curvature $R_\alpha{}^\beta$.

In this paper we continue a study of the 3D gravitational model of 
Mielke-Baekler which was initiated in \cite{cs}. Within the Poincar\'e gauge
approach, the Lagrangian of the model reads:
\begin{eqnarray}\label{mblag}
  V_{\rm MB} &=& -\,\frac{\chi}{2\ell}\,R^{\alpha\beta}\wedge\eta_{\alpha\beta}
-\frac{\Lambda}{\ell}\,\eta +\frac{\theta_{\rm T}}{2\ell^2}\,\vartheta^\alpha
 \wedge T_\alpha \nonumber\\
&& -\,\frac{\theta_{\rm L}}{2}\,\left(\Gamma_\alpha{}^\beta\wedge d
 \Gamma_\beta{}^\alpha - \frac{2}{3}\,\Gamma_\alpha{}^\beta\wedge 
 \Gamma_\beta{}^\gamma \wedge\Gamma_\gamma{}^\alpha \right) + L_{\rm mat}\,.
\end{eqnarray}
The coupling constants $\chi, \theta_{\rm T}, \theta_{\rm L}$ are
dimensionless, and they specify the usual Hilbert-Einstein, the
translational, and the rotational Chern-Simons terms, respectively.
The coupling constant $\ell$ has the dimension of length; $\Lambda$
is the cosmological constant. 

Our geometrical notations and conventions are as follows: $\eta$ is the
volume 3-form, $\eta_\alpha = e_\alpha\rfloor\eta$, $\eta_{\alpha\beta}
= e_\beta\rfloor\eta_\alpha$, and finally $\eta_{\alpha\beta\gamma} =
e_\gamma\rfloor\eta_{\alpha\beta}$ is the totally antisymmetric Levi-Civita 
tensor. As usual, Greek indices, $\alpha,\beta,\dots =0,1,2$, label the 
(co)frame components, whereas Latin indices, $i,j,\dots = 0,1,2$, label 
the components with respect to a local coordinate basis. 

In contrast to \cite{cs}, we consider not only a vacuum situation, but take 
the nontrivial matter source into account. Main attention is paid to the 
case when the dynamical spin current of matter is trivial, thus leaving the 
energy-momentum current as the only source of the gravitational field. 
In Sec.~\ref{EE}, we demonstrate that a generic model (\ref{mblag}) for
arbitrary coupling constants $\chi, \theta_{\rm T}, \theta_{\rm L}$ 
(except for some special choices) is equivalent to a 3D Einstein
gravitational theory with certain effective energy-momentum current. The
latter differs essentially from the original energy-momentum only when
$\theta_{\rm L}\neq 0$, otherwise the two currents are merely proportional
to each other. The exceptional models are, as a rule, inconsistent when 
matter is present, imposing unacceptable mathematical and physical constraints 
on the structure of the energy-momentum current. 

Sec.~\ref{PPsol} is devoted to the derivation of new solutions for a 
nontrivial electromagnetic source. Specifically, we consider the plane-fronted
gravitational wave configurations. In Sec.~\ref{HSsol} we analyze the case
of an ideal fluid source. In particular, the circularly symmetric solutions 
are derived here, some of which can be interpreted as black holes. As another 
class of exact solutions we obtain anisotropic cosmological models of a 
Heckmann-Sch\"ucking type. Finally, in Sec.~\ref{DC} we summarize the 
results obtained.

\section{Effective Einstein theory}\label{EE}

The gravitational field equations are derived from the variation of 
(\ref{mblag}) with respect to coframe and (Lorentz) connection:
\begin{eqnarray}
\label{field1}
\frac{\chi}{2} \, \eta_{\alpha\beta\gamma} \, R^{\beta\gamma} +
\Lambda \, \eta_{\alpha} - \frac{\theta_{{\rm T}}}{\ell} \,
T_\alpha &=& \ell \, \Sigma_\alpha\,,\\
\label{field2} \frac{\chi}{2} \, \eta_{\alpha\beta\gamma} \, T^{\gamma} -
\frac{\theta_{{\rm T}}}{2\ell} \, \vartheta_\alpha\wedge\vartheta_\beta -
\theta_{{\rm L}} \, \ell \, R_{\alpha\beta} &=& \ell \,
\tau_{\alpha\beta} \,.
\end{eqnarray}
The 2-forms of the material energy-momentum and spin currents are defined 
by the variational derivatives of the Lagrangian $L_{\rm mat}$ of matter
$\Sigma_\alpha:=\delta L_{\rm mat}/\delta \vartheta^\alpha$
and $\tau_{\alpha\beta}:=\delta L_{\rm mat}/\delta\Gamma_\alpha
{}^\beta$, respectively.

The algebraic system (\ref{field1})-(\ref{field2}) can be resolved
with respect to the torsion and the curvature 2-forms:
\begin{eqnarray}
T^\alpha &=& {\frac 2 \ell}\left({\cal T}\,\eta^\alpha + \alpha_{\rm L}\,
\ell^3\,\Sigma^\alpha + \beta\,\ell^2{\frac 12}\,\eta^{\alpha\beta\gamma}
\,\tau_{\beta\gamma}\right),\label{Ta}\\
R^{\alpha\beta} &=& {\frac 1 {\ell^2}}\left({\cal R}\,\vartheta^\alpha
\wedge\vartheta^\beta +  \alpha_{\rm T}\,\ell^2\,\tau^{\alpha\beta} + 
\beta\,\ell^3\,\eta^{\alpha\beta\gamma}\,\Sigma_\gamma\right).\label{Rab}
\end{eqnarray}
The numeric coefficients are constructed from the coupling constants of
the model: 
\begin{eqnarray}
{\cal T} &=& {\frac {-\frac{\theta_{\rm T}}2\,\chi +\Lambda
\ell^2 \theta_{\rm L}} {\chi^2 + 2\theta_{\rm T}\theta_{\rm L}}} \qquad 
{\cal R} = -\,{\frac {\theta_{\rm T}^2 + \chi\Lambda\ell^2} {\chi^2 + 
2\theta_{\rm T}\theta_{\rm L}} },\label{RT}\\ 
\alpha_{\rm L} &=& -\,{\frac {\theta_{\rm L}} {\chi^2 + 2\theta_{\rm T} 
\theta_{\rm L}}},\qquad \alpha_{\rm T} = -\,{\frac {2\theta_{\rm T}}
{\chi^2 + 2\theta_{\rm T}\theta_{\rm L}}},\qquad \beta = -\,{\frac \chi 
{\chi^2 + 2\theta_{\rm T}\theta_{\rm L}}}.\label{albe}
\end{eqnarray}

Since in our work we will be mainly interested in the electromagnetic field 
and the ideal (spinless) fluid as the material sources, we now will specialize 
to the case of the matter without dynamical spin, $\tau_{\alpha\beta} =0$. 
Then, as we know, the energy-momentum current is symmetric, i.e., 
\begin{equation}
e_\alpha\rfloor\Sigma^\alpha = {}^\ast(\vartheta^\alpha\wedge{}^\ast
\Sigma_\alpha) = 0.\label{symmetry}
\end{equation}
However, the trace of the energy-momentum
\begin{equation}
\Sigma = {}^\ast (\vartheta^\alpha\wedge\Sigma_\alpha) = e_\alpha\rfloor
\,{}^\ast\Sigma^\alpha\label{trace}
\end{equation}
is nontrivial, in general. 

We now demonstrate that the above system (\ref{field1})-(\ref{field2})
is equivalent to an effective Einstein equation. In order to derive this, 
we note that the local connection 1-form splits into a Riemannian and the
contortion parts:
\begin{equation}\label{Gamma}
  \Gamma_{\alpha\beta} = \widetilde{\Gamma}_{\alpha\beta} + 
  K_{\alpha\beta},\qquad K_{\alpha\beta} =  - e_{[\alpha}\rfloor T_{\beta]} 
 + \frac{1}{2} \, \left( e_\alpha \rfloor e_\beta \rfloor T_\gamma
  \right) \vartheta^\gamma .
\end{equation}
The curvature 2-form then decomposes accordingly:
\begin{equation}
R_\alpha{}^\beta = \widetilde{R}_\alpha{}^\beta + \widetilde{D}
K_\alpha{}^\beta - K_\alpha{}^\gamma\wedge K_\gamma{}^\beta\,.
\end{equation}
Hereafter we denote by the tilde the geometrical objects and operators
constructed with the help of the Riemannian connection. 
Using (\ref{field1}), we find
\begin{equation}
K_{\alpha\beta} = {\cal K}\,\eta_{\alpha\beta} + 2\alpha_{\rm L}\ell^2
\,\eta_{\alpha\beta\gamma}\,{}^\ast\Sigma^\gamma,\qquad {\cal K}:= 
{\cal T}/\ell - \alpha_{\rm L}\ell^2\,\Sigma.
\end{equation}
It is important to note that ${}^{\ast\ast}=-1$ for {\it all} forms in a
3-dimensional spacetime with the metric of Lorentzian signature. Then we have 
\begin{eqnarray}
\widetilde{D}K_\alpha{}^\beta &=& -\eta_\alpha{}^\beta\wedge d{\cal K} 
+ 2\alpha_{\rm L}\ell^2\,\eta_\alpha{}^{\beta\gamma}\,\widetilde{D}
\,{}^\ast\Sigma_\gamma,\\
- K_\alpha{}^\gamma\wedge K_\gamma{}^\beta &=& -({\cal T}^2/\ell^2 - 
\alpha_{\rm L}^2\ell^4\,\Sigma^2)\,\vartheta_\alpha
\wedge\vartheta^\beta + 2{\cal K}\alpha_{\rm L}\ell^2\,\eta_\alpha
{}^{\beta\gamma}\Sigma_\gamma - 4\alpha_{\rm L}^2\ell^4
\,{}^\ast\Sigma_\alpha\wedge{}^\ast\Sigma^\beta. 
\end{eqnarray}
In deriving these formulas we used the symmetry of the energy-momentum current.

As a result, each of the field equations (\ref{field1}) and (\ref{field2}) 
takes the form of the effective Einstein equation in three dimensions:
\begin{equation}
{\frac 12}\,\eta_{\alpha\beta\gamma}\,\widetilde{R}^{\beta\gamma} +
\Lambda^{\rm eff}\,\eta_\alpha = \ell\,\Sigma_\alpha^{\rm eff}.\label{eff}
\end{equation}
Here the effective cosmological term is defined by
\begin{equation}
\Lambda^{\rm eff} = -\,{\frac {{\cal R} + {\cal T}^2}{\ell^2}},\label{Leff}
\end{equation}
whereas the effective energy-momentum 2-form reads
\begin{eqnarray}
\Sigma_\alpha^{\rm eff} &=& (2\alpha_{\rm L}{\cal T} - \beta)\,\Sigma_\alpha 
+ \alpha_{\rm L}\ell\left(\vartheta_{\alpha}\wedge d\,\Sigma
+ 2\widetilde{D}\,{}^\ast\Sigma_\alpha\right)\nonumber\\ 
&& + \,\alpha_{\rm L}^2\ell^3\left( - \Sigma^2\,\eta_\alpha -2\Sigma
\,\Sigma_\alpha + 2\eta_{\alpha\beta\gamma}\,{}^\ast\Sigma^\beta\wedge
{}^\ast\Sigma^\gamma\right).\label{Teff}
\end{eqnarray}
As an important consistency check, we need to verify that the effective 
energy-momentum is conserved, just like the original current $\widetilde{D}
\Sigma_\alpha =0$. Taking the covariant derivative, we find 
\begin{equation}
\widetilde{D}\Sigma_\alpha^{\rm eff} =  -\,2\alpha_{\rm L}\ell\left[
\widetilde{R}_\alpha{}^\beta\wedge{}^\ast\Sigma^\beta + \alpha_{\rm L}\ell^2
\,d\Sigma\wedge (\Sigma\eta_\alpha + \Sigma_\alpha) + 2\alpha_{\rm L}\ell^2
\,\eta_{\alpha\beta\gamma}\,{}^\ast\Sigma^\beta\wedge \widetilde{D}
\,{}^\ast\Sigma^\gamma\right].
\end{equation}
Here we used the Ricci identity $\widetilde{D}\widetilde{D}\,{}^\ast
\Sigma_\alpha = -\,\widetilde{R}_\alpha{}^\beta\wedge{}^\ast\Sigma_\beta$.
It remains now to substitute the curvature 2-form from the field equation
(\ref{eff}) and to use (\ref{Teff}). The symmetry (\ref{symmetry}) of the 
original energy-momentum yields $\eta_{[\alpha}\wedge{}^\ast\Sigma_{\beta]} 
= 0$ and $\Sigma_{[\alpha}\wedge{}^\ast\Sigma_{\beta]} =0$. Using all this
together with the identity $\Sigma\eta_\alpha + \Sigma_\alpha = -\,\eta_{\alpha
\beta}\wedge{}^\ast\Sigma^\beta$, we finally obtain the conservation law
\begin{equation}
\widetilde{D}\Sigma_\alpha^{\rm eff} = 0.
\end{equation}

As another consistency check, let us consider the limiting case of 
$\theta_{\rm T} = \theta_{\rm L} =0$. Then (\ref{RT}) and (\ref{Leff}) 
yield $\Lambda^{\rm eff} = \Lambda/\chi$, while (\ref{Teff}) reduces to 
$\Sigma_\alpha^{\rm eff} = \Sigma_\alpha/\chi$. We thus recover the the 
correct Einstein theory in three dimensions (or, equivalently, the 
model of Witten \cite{witten}). 

A very close model arises when $\theta_{\rm L} = 0$. Then we find 
$\Lambda^{\rm eff} = \Lambda/\chi + 3\theta_{\rm T}^2/(4\ell^2\chi^2)$, and
again $\Sigma_\alpha^{\rm eff} = \Sigma_\alpha/\chi$. In other words, when 
the Hilbert-Einstein Lagrangian is modified only by the translational 
Chern-Simons term, the resulting dynamics differs from the Einstein 
theory only via the shifted cosmological constant. 

In contrast, the case $\theta_{\rm T} = 0$ represents a nontrivial extension of
the Einstein theory. The cosmological constant and the Einstein gravitational
coupling constant then are replaced by $(\Lambda/\chi)\rightarrow(\Lambda/\chi)
(1 - \Lambda\ell^2\theta_{\rm L}^2/\chi^3)$ and $(1/\chi)\rightarrow (1/\chi)
(1 - 2\Lambda\ell^2\theta_{\rm L}^2/\chi^3)$, respectively. In addition, the
effective energy-momentum (\ref{Teff}) contains the new terms proportional 
to $\theta_{\rm L} = 0$. 

As we see from (\ref{RT})-(\ref{albe}), the reduction to the effective 
Einstein theory takes place for generic models with $\chi^2 + 2\theta_{\rm T}
\theta_{\rm L}\neq 0$. An exceptional case $\chi^2 + 2\theta_{\rm T}
\theta_{\rm L} = 0$ should be analyzed separately. Assuming all the coupling 
constants to be nonzero, we then find that, for example, the translational 
coupling can be expressed in terms of the Lorentzian one, $\theta_{\rm T} =
- \chi^2/(2\theta_{\rm L})$. This immediately imposes the algebraic 
consistency condition on the right-hand sides of the field equations
(\ref{field1}) and (\ref{field2}): $\ell\Sigma_\alpha = (\chi/\theta_{\rm L})
\,\eta_{\alpha\beta\gamma}\,\tau^{\beta\gamma} + \eta_\alpha\,[\Lambda -
\chi^3/(4\ell^2\theta_{\rm L}^2)]$. In particular, the energy-momentum turns
out to be proportional to a cosmological term for a matter with the vanishing 
dynamical spin. The exceptional case includes both models with Lagrangians
containing just one Chern-Simons term. For the purely translational 
Chern-Simons gravity with $\theta_{\rm T}\neq 0$ and $\chi=0,\,\theta_{\rm L} 
= 0$, the spin must be constant $\tau_{\alpha\beta} = - (\theta_{\rm T}/2
\ell^2)\vartheta_\alpha\wedge\vartheta_\beta$ and the spinless matter case 
is ruled out as inconsistent. Analogously, the purely Lorentzian Chern-Simons 
gravity with $\theta_{\rm L}\neq 0$ and $\chi=0,\,\theta_{\rm T} = 0$ is
inconsistent for any nontrivial energy-momentum except the cosmological
term $\ell\Sigma_\alpha = \Lambda\eta_\alpha$.

\section{Exact solutions with electromagnetic source}\label{PPsol}

As a first new solution, we derive the $pp$-wave solution in 3D gravity.
Plane-fronted gravitational wave solutions represent an important class
of spacetimes in 4 dimensions \cite{pp4}, as well as in higher dimensions
\cite{ppN}. A particular interest to the $pp$-waves is related with their
role in the string theory \cite{ppS}. Recently, the wave solutions were also
studied in the metric-affine gravity models \cite{ppRC}.

\subsection{Geometry of a $pp$-wave}

We choose the local coordinates $(\sigma, \rho, z)$, and take the line 
element in the form
\begin{equation}
ds^2 = g_{\alpha\beta}\,\vartheta^\alpha\otimes\vartheta^\beta,\label{ds2}
\end{equation}
with the half-null Lorentz metric 
\begin{equation}
g_{\alpha\beta} = \left(\begin{array}{ccc}0&1&0\\ 1&0&0\\ 0&0&1\end{array}
\right).\label{met}
\end{equation}
The components of the coframe 1-form are given by
\begin{equation}
\vartheta^{\widehat{0}} = -\,d\sigma,\qquad \vartheta^{\widehat{1}} 
= \left({\frac q p}\right)^2\left(s\,d\sigma + d\rho\right), 
\qquad \vartheta^{\widehat{2}} = {\frac 1 p}\,dz.\label{cof}
\end{equation}
The dual frame basis (such that $e_\alpha\rfloor\vartheta^\beta = 
\delta_\alpha^\beta$) is easily constructed:
\begin{equation}
e_{\widehat{0}} = -\,\partial_\sigma + s\,\partial_\rho,\qquad 
e_{\widehat{1}} = \left({\frac p q}\right)^2\partial_\rho,\qquad
e_{\widehat{2}} = p\,\partial_z. 
\end{equation}
We choose the functions $p(z), q(z), s(\sigma,\rho, z)$ as follows:
\begin{equation}
p = 1 + {\frac \lambda 4}\,z^2,\qquad q = 1 - {\frac \lambda 4}
\,z^2,\qquad s = -\,{\frac \lambda 2}\,\rho^2 + {\frac {\sqrt{p}}
{2\,q}}\,H(\sigma, z).\label{pqs}
\end{equation}
This ansatz should be compared to the four-dimensional case \cite{pleb}.
Then, the Riemannian curvature 2-form reads:
\begin{equation}\label{curv}
\widetilde{R}^{\alpha\beta}= -\,\lambda\,\vartheta^\alpha\wedge\vartheta^\beta 
- \eta^{\alpha\beta\gamma}k_\gamma k^\lambda\eta_\lambda\,p(q^2s'/p)'.
\end{equation}
Here we denote the null vector $k_\alpha = \delta_{\alpha}^{\widehat{0}}$ 
with $k_\alpha k^\alpha = 0$. The derivative with respect to the $z$ 
coordinate is denoted by the prime $(')$. 

The resulting geometry is regular in sense that the polynomial curvature 
invariants are always constant, irrespectively of the form of the function 
$H(\sigma, z)$. For example, the quadratic invariant of the Riemannian 
curvature reads
\begin{equation}
\widetilde{R}^{\alpha\beta}\wedge{}^\ast\widetilde{R}_{\alpha\beta} 
= 6\lambda^2\eta.\label{R2}
\end{equation}
The cubic invariants are proportional to $\lambda^3$, and so on.

\subsection{Electromagnetic source}

Let the material source be represented by the
energy-momentum of an electromagnetic wave. Taking the potential 1-form 
\begin{equation}
A = \varphi(\sigma,z)\,\vartheta^{\widehat{0}},
\end{equation}
we find the electromagnetic field 2-form $F =dA = {\frac 12}\,F_{\alpha\beta}
\vartheta^\alpha\wedge\vartheta^\beta$ with the tensor components 
\begin{equation}
F_{\alpha\beta} = 2n_{[\alpha}\,k_{\beta]}.\label{F}
\end{equation}
Here $n^\alpha\,k_\alpha =0$. In terms of the vector potential, the components
of the covector $n$ read $n_\alpha = \delta_\alpha^2\,p\varphi'$. The unknown 
scalar function $\varphi$ is determined by the Maxwell's equation 
$d\,{}^\ast\! F = 0$ which for the metric (\ref{ds2})-(\ref{cof}) 
reduces to the partial differential equation
\begin{equation}
(p\varphi')' = 0.\label{ddf}
\end{equation}
This equation is easily integrated, yielding $\varphi' = \nu(\sigma)/p$
with an arbitrary function $\nu(\sigma)$. The electromagnetic energy-momentum 
current 2-form then reads:
\begin{equation}\label{sigM}
\Sigma_\alpha = {\frac 12}\left[(e_\alpha\rfloor F)\wedge{}^\ast\!F - F\,(
e_\alpha\rfloor\,{}^\ast\!F)\right] = \nu^2\,k_\alpha k^\beta\,\eta_{\beta}.
\end{equation}
Since the vector $k_\alpha$ is null, we find $\vartheta^\alpha\wedge 
\Sigma_\alpha =0$. Furthermore, for the same reason it is obvious that 
\begin{equation}
\Sigma_\alpha\wedge\Sigma_\beta = 0,\qquad \Sigma_\alpha\wedge{}^\ast
\Sigma_\beta = 0,\qquad {}^\ast\Sigma_\alpha\wedge{}^\ast\Sigma_\beta = 0.
\end{equation} 
Hence $\eta_{\alpha\beta\gamma}{}^\ast\Sigma^\beta\wedge{}^\ast\Sigma^\gamma 
= 0$. Finally, the direct computation yields 
\begin{equation}\label{dsigM}
\widetilde{D}\,{}^\ast\Sigma_\alpha = -\,{\frac {\lambda z}q}\,\Sigma_\alpha. 
\end{equation}

\subsection{Explicit $pp$-wave solutions}

We now can substitute (\ref{curv}) and the energy-momentum (\ref{sigM}) and 
(\ref{dsigM}) into the effective Einstein equation (\ref{eff}). The latter 
yields the algebraic relation $\lambda = \Lambda^{\rm eff}$ and the 
differential equation:
\begin{equation}\label{dds}
\left[{\frac {q^2}p}\left({\frac {\sqrt{p}}{2q}}H\right)'\right]' 
= \nu^2\ell\left({\frac {2\alpha_{\rm L}{\cal T} - \beta} p} 
- {\frac {2\alpha_{\rm L}\ell\,\lambda z}{pq}}\right). 
\end{equation}
Integration is straightforward and yields the general solution:
\begin{eqnarray}
H(\sigma, z) &=& {\frac 1 {\sqrt{p}}}\,\Big\{\mu_1(\sigma)\,q + \mu_2(\sigma)
\,z + 2\nu^2\ell\left[(2\alpha_{\rm L}{\cal T} - \beta)\,z - 2\alpha_{\rm L}
\ell\,q\right](2/\sqrt{\lambda})\,\arctan(\sqrt{\lambda}z/2) \nonumber\\
&&\qquad\qquad -\,2\nu^2\ell\left[ 2\alpha_{\rm L}\ell\,z  + (2\alpha_{\rm L}
{\cal T} - \beta)\,q/\lambda\right]\,\ln(|p/q|)\Big\}.\label{solution1}
\end{eqnarray}
With the two arbitrary functions $\mu_{1,2}(\sigma)$, the first two terms
above represent the general solution of the homogeneous equation with the
vanishing right-hand side in (\ref{dds}). 

The resulting configuration turns out to be localized along the $z$-coordinate.
Namely, for large positive and negative values of $z\rightarrow\pm\infty$, 
the metric function reads $s = -\,\lambda\rho^2/2 + \nu_1/2 + {\cal O}(1/z)$, 
with $\nu_1 = \mu_1 \mp 4\pi\nu^2\alpha_{\rm L}\ell^2/\sqrt{\lambda}$. The 
geometry is thus asymptotically de Sitter in these limits. The function $s$ 
blows up when $q=0$, i.e., at $z = \pm 2/\sqrt{\lambda}$, however there is 
no curvature singularity at these points, cf. (\ref{R2}). All this refers
to the case when the effective cosmological constant is positive, 
$\Lambda^{\rm eff} > 0$. 

When $\Lambda^{\rm eff} < 0$, we obtain a different solution with the 
negative $\lambda = -|\Lambda^{\rm eff}|$:
\begin{eqnarray}
H(\sigma, z) &=& {\frac 1 {\sqrt{p}}}\,\Big\{\mu_1(\sigma)\,q + \mu_2(\sigma)
\,z + 2(\nu^2\ell/\sqrt{|\lambda|})\left[(2\alpha_{\rm L}{\cal T}\! - \!\beta)
\,z - 2\alpha_{\rm L}\ell\,q\right]\,\ln[(q + \sqrt{|\lambda|}z)/p]
\nonumber\\
&&\qquad\qquad -\,2\nu^2\ell\left[ 2\alpha_{\rm L}\ell\,z  + (2\alpha_{\rm L}
{\cal T} - \beta)\,q/\lambda\right]\,\ln(|p/q|)\Big\},\label{solution2}
\end{eqnarray}
where $\mu_{1,2}(\sigma)$ are again two arbitrary functions. The properties
of this solution are rather similar to those of (\ref{solution1}). Here we
again observe the similar asymptotic behavior $s = -\,\lambda\rho^2/2 + 
\mu_1/2 + {\cal O}(1/z)$ for $z\rightarrow\pm\infty$. And analogously,
the function $s$ diverges at $z = \pm 2/\sqrt{|\lambda|}$, although the 
curvature invariants are everywhere regular. 

Finally, for the vanishing effective cosmological constant $\Lambda^{\rm eff}
= 0$ we find $\lambda =0$, hence $p=q=1$, and the integration of (\ref{dds})
yields
\begin{equation}
H(\sigma, z) = \mu_1(\sigma) + \mu_2(\sigma)\,z + \nu^2\ell\,(2\alpha_{\rm L}
{\cal T} - \beta)\,z^2. 
\end{equation}
In this case all the curvature invariants (algebraic and differential) are
trivial.

\section{Exact solutions with an ideal fluid}\label{HSsol}

Let us consider now the ideal fluid which is characterized by the energy
density $\varepsilon$, pressure $p$ and the 2-form of the flow $u$. The
energy-momentum current then reads
\begin{equation}\label{fluid}
\Sigma_\alpha = p\,\eta_\alpha - (\varepsilon + p)u\,e_\alpha\rfloor{}^\ast u.
\end{equation}
As usual, we impose the normalization condition $u\wedge{}^\ast u = \eta$. 
It seems worthwhile to note that the covector of velocity is defined by the
Hodge dual, $-{}^\ast u$, of the flow 2-form $u$. The conservation law 
$\widetilde{D}\Sigma_\alpha = 0$ comprises the two well-known balance 
relations for the energy density and the pressure:
\begin{eqnarray}
u\wedge d\varepsilon + (\varepsilon + p)\,du &=& 0,\label{de}\\
dp + {}^\ast u\,{}^\ast(u\wedge dp) + a &=& 0.\label{dp}
\end{eqnarray}
Here the fluid acceleration 1-form is defined by $a:= {}^\ast(u\wedge 
\widetilde{D}e_\alpha\rfloor{}^\ast u)\,\vartheta^\alpha$. It evidently
satisfies the ``orthogonality" condition $u\wedge a = 0$. 

Substituting (\ref{fluid}) into (\ref{Teff}), we find for the last term of
the effective energy-momentum: 
\begin{equation}
- \Sigma^2\,\eta_\alpha -2\Sigma\,\Sigma_\alpha + 2\eta_{\alpha\beta\gamma}
\,{}^\ast\Sigma^\beta\wedge{}^\ast\Sigma^\gamma = -\,2\varepsilon
\Sigma_\alpha - \varepsilon^2\eta_\alpha. 
\end{equation}

\subsection{Stationary rotating configurations}\label{HSsol1}

Let us impose the differential condition on the fluid flow:
\begin{equation}
\widetilde{D}e_\alpha\rfloor{}^\ast u = \mu\,e_\alpha\rfloor u.\label{du1}
\end{equation}
Here $\mu$ is an arbitrary constant. Multiplying (\ref{du1}) from the left 
by the coframe $\vartheta^\alpha$, we find 
\begin{equation}
d\,{}^\ast\!u = -\,2\mu\,u.\label{du2}
\end{equation}
The above conditions essentially constrain the kinematics of the fluid.
Directly from (\ref{du1}) we derive that the acceleration vanishes, $a=0$,
and hence (\ref{du2}) shows that the vorticity is nontrivial, $\omega =
d\,{}^\ast\!u = -\,2\mu\,u$. On the other hand, taking the exterior 
differential of (\ref{du2}), we find that the volume expansion is zero, $du 
=0$. In other words, we have a class of stationary rotating configurations.

Using these kinematic properties in the conservation law (\ref{de})-(\ref{dp}),
we conclude that the constant energy density and pressure are compatible with
the above assumptions:
\begin{equation}
\varepsilon = \varepsilon_0,\qquad p = p_0.\label{ep0}
\end{equation}
Using then (\ref{du1})-(\ref{ep0}) in (\ref{eff}), we obtain the Einstein
field equation
\begin{equation}
{\frac 12}\,\eta_{\alpha\beta\gamma}\,\widetilde{R}^{\beta\gamma} +
\underline{\Lambda}\,\eta_\alpha = \ell\,\underline{\Sigma}_\alpha,\label{eff2}
\end{equation}
for the effective fluid energy-momentum
\begin{equation}
\underline{\Sigma}_\alpha = \underline{p}\,\eta_\alpha - 
(\underline{\varepsilon} + \underline{p})u\,e_\alpha\rfloor{}^\ast u
\end{equation}
with the constant energy density and pressure defined by
\begin{eqnarray}
\underline{\varepsilon} &=&(2\alpha_{\rm L}{\cal T} - \beta - 6\mu
\alpha_{\rm L}\ell - 2\alpha_{\rm L}^2\ell^3\varepsilon_0)\,\varepsilon_0,\\
\underline{p} &=& (2\alpha_{\rm L}{\cal T} - \beta - 6\mu\alpha_{\rm L}\ell
- 2\alpha_{\rm L}^2\ell^3\varepsilon_0)\,p_0.
\end{eqnarray}
The shifted ``cosmological constant" reads:
\begin{equation}
\underline{\Lambda} = \Lambda^{\rm eff} + \alpha_{\rm L}\ell\left[2\mu
(\varepsilon_0 - 2p_0) + \alpha_{\rm L}\ell^2\varepsilon_0^2\right]. 
\end{equation}

As an example, let us derive the black hole type solution. We choose the
local coordinates $(t, r, \phi)$ and write the line element $ds^2 = 
g_{\alpha\beta}\,\vartheta^\alpha\otimes\vartheta^\beta$ with the Minkowski
flat metric $g_{\alpha\beta} = {\rm diag}(-1,+1,+1)$ and the coframe
\begin{equation}
\vartheta^{\widehat{0}} = dt - h(r)\,d\phi,\qquad 
\vartheta^{\widehat{1}} = {\frac 1 {f(r)}}\,dr,\qquad 
\vartheta^{\widehat{2}} = f(r)\,rd\phi.\label{cofBH}
\end{equation}
For the flow 2-form of the perfect fluid we assume the ansatz
\begin{equation}
u = \vartheta^{\widehat{1}}\wedge\vartheta^{\widehat{2}}.
\end{equation}
Then from the differential condition (\ref{du1}), (\ref{du2}) we find
\begin{equation}
{\frac 1 r}\,{\frac {dh}{dr}} = 2\mu,\qquad {\rm or}
\qquad h(r) = \mu\,r^2 + h_0,\label{hsol}
\end{equation}
with some constant $h_0$. Substituting (\ref{cofBH}) into the effective 
Einstein equation (\ref{eff2}), we find the second unknown function
\begin{equation}
f^2 =  b_0 + b_1/r^2 + k\,r^2.\label{fsol}
\end{equation}
The constant energy density $\varepsilon_0$ turns out to be an essential
parameter which determines the other geometric and physical quantities of the
solution. The constants $b_0, b_1$ are arbitrary, whereas $k$ and the pressure
$p_0$ are given by
\begin{eqnarray}
k &=& {\frac {2\alpha_{\rm L}\ell\left[\Lambda^{\rm eff}(3\mu + 
\hat{\varepsilon}) - (\mu + \hat{\varepsilon})^3\right] + (2\alpha_{\rm L}
{\cal T}\! - \!\beta)\left[3(\mu + \hat{\varepsilon})^2 - \Lambda^{\rm eff}
\right] - \ell\varepsilon_0(2\alpha_{\rm L}{\cal T}\! - \!\beta)^2}
{4\left[2\alpha_{\rm L}{\cal T} - \beta - 2\alpha_{\rm L}\ell (\mu + 
\hat{\varepsilon})\right]}},\\
p_0 &=& {\frac {\mu^2 + \Lambda^{\rm eff} + \hat{\varepsilon}(2\mu + 
\hat{\varepsilon})} {\ell\left[2\alpha_{\rm L}{\cal T} - \beta - 
2\alpha_{\rm L}\ell (\mu + \hat{\varepsilon})\right]}}.
\end{eqnarray}
Here we denoted $\hat{\varepsilon} := \alpha_{\rm L}\ell^2\varepsilon_0$.
In the limit of the Einstein theory with both Chern-Simons terms absent,
$\theta_{\rm L} = \theta_{\rm T} = 0$, the above quantities reduce to
\begin{equation}
k = {\frac {3\chi\mu^2 - \Lambda - \ell\varepsilon_0} {4\chi}},\qquad 
\ell p_0 = \chi\mu^2 + \Lambda. 
\end{equation} 

In general, after a convenient rescaling of the time and the angular 
coordinates, 
we can write the above solution as
\begin{equation}\label{metBH}
ds^2 = - N^2\,dt^2 + {\frac {dr^2}{f^2}} + g_{22}\,(d\phi + N^\phi\,dt)^2,
\end{equation}
where (with the above constant parameters fixed as $b_0 = - m, b_1 = J^2/4,
h_0 = J/2$)
\begin{eqnarray}
f^2 = - m + (J/2r)^2 + k\,r^2,\qquad N^2 = f^2\left(1 + {\frac {r^2(k - \mu^2)}
{\mu J - m}}\right)^{-1},\\
g_{22} = r^2\left(1 + {\frac {r^2(k - \mu^2)}{\mu J - m}}\right),\qquad
N^\phi = \left(\mu - {\frac J {2r^2}}\right)\left(1 + {\frac {r^2(k - \mu^2)}
{\mu J - m}}\right)^{-1}. 
\end{eqnarray}
The metric (\ref{metBH}) describes a black hole type configurations. The
horizons are determined by
\begin{equation}
r_h^2 = {\frac {m\pm \sqrt{m^2 - J^2k}}{2k}}.
\end{equation}
Strictly speaking, this solution is no black hole because both the quasilocal 
angular momentum and the quasilocal mass \cite{brown,der} can be (one or both) 
divergent, in general. Accordingly, no definite finite mass and spin then can 
be attributed to such a configuration. In particular, let us compute the 
quasilocal angular momentum:
\begin{equation}
j(r) = {\frac {f(g_{22})^{3/2}}N}{\frac {dN^\phi}{\!dr}} = J + {\frac 
{2r^2(\mu^2 - k)(\mu^2r^2 - J)} {\mu J - m}}.
\end{equation}
This quantity obviously diverges for $r\rightarrow\infty$. The quasilocal
mass is given by a more complicated expression:
\begin{equation}
m(r) = {\frac {\left(m + {\cal O}(1/r^2)\right)\left(1 + {\frac {2r^2(k - 
\mu^2)}{\mu J - m}}\right) - j(r)\left(\mu - {\frac J {2r^2}}\right)}
{1 + {\frac {r^2(k - \mu^2)}{\mu J - m}}}}.
\end{equation}
Because of the last term in the numerator, this quantity also diverges for 
$r\rightarrow\infty$. There are, however, some particular cases when both 
quasilocal quantities yield the finite limits at the spatial infinity. The 
first case takes place for $\mu = 0$. Then we find the finite mass of the 
solution, $M = m(\infty) = 2m$. The quasilocal angular momentum is, however, 
still divergent unless $J = 0$. The resulting non-rotating configuration 
describes a nontrivial black hole with a horizon at $r_h^2 = m/k$. The second 
case happens when $k = \mu^2$. Then the resulting mass and spin are equal 
$M = m$ and $j(\infty) = J$, respectively. In this case we recover the BTZ 
solution \cite{btz} which describes a black hole when $m \geq \mu J$, and 
a naked singularity for $\mu J > m$. It is instructive to compare this with 
the solution \cite{gurses} in the topologically massive (DJT) gravity in 
three dimensions \cite{djt}. 

The curvature of the above solution is completely regular. We find for the
anholonomic components of the Riemannian curvature 2-form
\begin{equation}
\widetilde{R}^{\widehat{0}\widehat{1}} = \mu^2\,\vartheta^{\widehat{0}}\wedge
\vartheta^{\widehat{1}},\quad \widetilde{R}^{\widehat{0}\widehat{2}} = \mu^2
\,\vartheta^{\widehat{0}}\wedge\vartheta^{\widehat{2}},\quad
\widetilde{R}^{\widehat{1}\widehat{2}} = (4k - 3\mu^2)\,\vartheta^{\widehat{1}}
\wedge\vartheta^{\widehat{2}}.
\end{equation}
For $k = \mu^2$ we recover the curvature of the anti-de Sitter spacetime, in
complete agreement with the BTZ-limit.

\subsection{Cosmology with an ideal fluid}\label{HSsol2}

Let us analyze now the cosmological solutions. We choose the local 
coordinates $(t, x, y)$ and write the line element $ds^2 = g_{\alpha\beta}
\,\vartheta^\alpha\otimes\vartheta^\beta$ with the Minkowski flat metric 
$g_{\alpha\beta} = {\rm diag}(-1,+1,+1)$ and the coframe
\begin{equation}
\vartheta^{\widehat{0}} = dt,\qquad \vartheta^{\widehat{1}} = R\,e^rdx,
\qquad \vartheta^{\widehat{2}} = R\,e^{-r}dy.\label{cofco}
\end{equation}
The two functions of the cosmological time, $R=R(t)$ and $r=r(t)$ describe
the evolution of an anisotropic 3D universe. Accordingly, we assume that
the energy density and the pressure depend on the time only: $\varepsilon =
\varepsilon(t)$, $p = p(t)$. Finally, we choose for the flow 2-form the 
ansatz $u = \vartheta^{\widehat{1}}\wedge\vartheta^{\widehat{2}}$, as above.
The energy conservation (\ref{de}) then yields the familiar equation:
\begin{equation}
\dot{\varepsilon} + 2{\frac {\dot{R}} R}\,(\varepsilon + p) = 0.\label{dec}
\end{equation}
The dot denotes the time derivative in this subsection. Choosing now the 
equation of state $p = \gamma\varepsilon$ (with constant $\gamma$),
we can integrate the above equation. This yields
\begin{equation}
\varepsilon = {\frac {\varepsilon_0}{R^{2(1 + \gamma)}}}.\label{eg}
\end{equation}
Substituting (\ref{fluid}) and (\ref{cofco}) into the effective Einstein 
equation (\ref{eff}), (\ref{Teff}), we find the following system of ordinary
differential equations:
\begin{eqnarray}
- {\frac {\dot{R}^2} {R^2}} + \dot{r}^2 + \Lambda^{\rm eff} &=& -\,\ell(
2\alpha_{\rm L}{\cal T} - \beta)\,\varepsilon + \alpha_{\rm L}^2\ell^4
\,\varepsilon^2,\label{eq0}\\
- {\frac {\ddot{R}} R} + 2{\frac {\dot{R}} R}\dot{r} + \ddot{r} - \dot{r}^2
+ \Lambda^{\rm eff} &=& \ell(2\alpha_{\rm L}{\cal T} - \beta)\,p 
- \alpha_{\rm L}^2\ell^4\,\varepsilon (\varepsilon + 2p),\label{eq1}\\
- {\frac {\ddot{R}} R} - 2{\frac {\dot{R}} R}\dot{r} - \ddot{r} - \dot{r}^2
+ \Lambda^{\rm eff} &=& \ell(2\alpha_{\rm L}{\cal T} - \beta)\,p 
- \alpha_{\rm L}^2\ell^4\,\varepsilon (\varepsilon + 2p),\label{eq2}\\
2\alpha_{\rm L}\ell\,\dot{r}\,(\varepsilon + p) &=& 0.\label{eq3}
\end{eqnarray}
Comparing (\ref{eq1}) and (\ref{eq2}), we find
\begin{equation}
\dot{r} = {\frac {c_0} {R^2}},\label{dr}
\end{equation}
with an arbitrary integration constant $c_0$. Moreover, as we can immediately
see, both equations (\ref{eq1}) and (\ref{eq2}) are redundant: They follow
from (\ref{eq0}) and the energy conservation (\ref{dec}). The further analysis
is based on the simple observation that the equation (\ref{eq3}) allows for
only one of the following three cases: (i) $\dot{r}=0$, (ii) $\varepsilon =
- p$, or (iii) $\alpha_{\rm L} = 0$. Let us consider these possibilities
separately.

\subsubsection{Case ($\dot{r}=0$): isotropic cosmology}

When $\dot{r}=0$, we can put $r = 0$ without a loss of generality. Then 
using (\ref{eg}) in (\ref{eq0}), we can integrate the latter equation for
an arbitrary value of $\gamma$. The form of the solution depends on the
sign of $\Lambda^{\rm eff}$ and on 
\begin{equation}
\Delta = {\frac {-\,\ell^2\varepsilon_0^2}{\chi^2 + 2\theta_{\rm T}
\theta_{\rm L}}}.
\end{equation}

Let us at first assume that $\lambda = \Lambda^{\rm eff} > 0$. Then for
the different values of $\Delta$ we find
\begin{eqnarray}
R^{2(1 + \gamma)} &=& {\frac 1 {2\lambda}}\left(- b + \sqrt{\Delta}\,\sinh[
2\sqrt{\lambda}(1+\gamma)(t - t_0)]\right),\qquad \Delta > 0,\label{del1}\\
R^{2(1 + \gamma)} &=& {\frac 1 {2\lambda}}\left(- b + \sqrt{|\Delta|}\,\cosh[
2\sqrt{\lambda}(1+\gamma)(t - t_0)]\right),\qquad \Delta < 0,\label{del2}\\
R^{2(1 + \gamma)} &=& {\frac 1 {2\lambda}}\left(- b + \,\exp[2\sqrt{\lambda}
(1+\gamma)(t - t_0)]\right),\qquad \Delta = 0.\label{del0}
\end{eqnarray}
Here $t_0$ is an arbitrary integration constant and $b := \ell(
2\alpha_{\rm L}{\cal T} - \beta)\,\varepsilon_0$. The last case, (\ref{del0}),
obviously describes the vacuum solution. The corresponding spacetime is a
3D de Sitter manifold. The solutions (\ref{del1}) and (\ref{del2}) describe
asymptotically (when $t\rightarrow\pm\infty$) de Sitter spacetimes. 

For the negative effective cosmological constant, $\lambda = - 
|\Lambda^{\rm eff}| < 0$, solution exists only for $\Delta < 0$ and reads:
\begin{equation}
R^{2(1 + \gamma)} = {\frac 1 {2|\lambda|}}\left(b + \sqrt{|\Delta|}\,\sin[
2\sqrt{|\lambda|}(1+\gamma)(t - t_0)]\right). 
\end{equation}

Finally, for the vanishing effective cosmological constant $\Lambda^{\rm eff}
=0$, we find the solution
\begin{equation}
R^{2(1 + \gamma)} = \alpha_{\rm L}^2\ell^4\varepsilon_0^2/b +
b\,(1 + \gamma)^2\,(t - t_0)^2. 
\end{equation}

\subsubsection{Case ($\varepsilon = -p$): anisotropic de Sitter}

For the vacuum equation of state we have $\gamma =-1$ and (\ref{eg}) yields 
constant energy density $\varepsilon = \varepsilon_0$. Using this, and
substituting (\ref{dr}) into (\ref{eq0}), we can integrate the resulting
equation. This yields
\begin{equation}\label{anids}
R^2 = {\frac {c_0}{\sqrt{Q}}}\,\sinh\left[2\sqrt{Q}(t - t_0)\right].
\end{equation}
Here $t_0$ is again an integration constant, and we denoted
\begin{equation}
Q := \Lambda^{\rm eff} + \ell(2\alpha_{\rm L}{\cal T} - \beta)\,\varepsilon_0 
- \alpha_{\rm L}^2\ell^4\varepsilon_0^2.\label{Qdef}
\end{equation}
Using (\ref{anids}) in (\ref{dr}), we find the second unknown function
\begin{equation}
r = {\frac 12}\,\ln\,\tanh\left[\sqrt{Q}(t - t_0)\right],
\end{equation}
and thus finally the line elements reads
\begin{equation}
ds^2 = -\,dt^2 + \sinh^2\left[\sqrt{Q}(t - t_0)\right]dx^2 + 
\cosh^2\left[\sqrt{Q}(t - t_0)\right]dy^2.
\end{equation}
The resulting geometry has the Riemannian curvature of the de Sitter 3D 
spacetime $\widetilde{R}^{\alpha\beta} = -\,Q\,\vartheta^\alpha\wedge
\vartheta^\beta$. The above was derived under the assumption that $Q>0$.
In case it happens that $Q < 0$, we find instead of (\ref{anids}):
\begin{equation}
R^2 = {\frac {c_0}{\sqrt{|Q|}}}\,\sin\left[2\sqrt{|Q|}(t - t_0)\right],
\end{equation}
and accordingly the line element is changed to
\begin{equation}
ds^2 = -\,dt^2 + \sin^2\left[\sqrt{|Q|}(t - t_0)\right]dx^2 + 
\cos^2\left[\sqrt{|Q|}(t - t_0)\right]dy^2.
\end{equation}
This metric describes the anti-de Sitter spacetime with the curvature
$\widetilde{R}^{\alpha\beta} = |Q|\,\vartheta^\alpha\wedge\vartheta^\beta$.

Finally, in case when $Q = 0$, we find $R^2 = 2c_0(t - t_0)$, and $2r 
= \ln(t - t_0)$ which yields the flat spacetime geometry
$ds^2 = -\,dt^2 + (t - t_0)^2\,dx^2 + dy^2$.

\subsubsection{Case ($\alpha_{\rm L} = 0$): Heckmann-Sch\"ucking type 
solutions}

As we saw in Sec.~\ref{EE}, when $\alpha_{\rm L} = \theta_{\rm L} = 0$, our 
effective theory 
in fact reduces to the Einstein gravity in three dimensions. This particular
case yields additional solutions to the system (\ref{eq0})-(\ref{eq3}) which
are analogs of the generalized Heckmann-Sch\"ucking cosmological solutions
obtained in four dimensions \cite{hs,kam}. As in \cite{kam}, we choose the
matter as the mixture of the three media - vacuum fluid, dust, and the
stiff matter. Accordingly, the energy density then reads as the sum of
the three terms, representing vacuum, dust and stiff matter, respectively:
\begin{equation}
\varepsilon = \varepsilon_0 + {\frac {\varepsilon_1}{R^2}} + {\frac
{\varepsilon_2}{R^4}},
\end{equation}
with the positive integration constants $\varepsilon_0, \varepsilon_1,
\varepsilon_2$. After substituting this into (\ref{eq0}), the integration
of (\ref{eq0}) and (\ref{dr}) is straightforward. The result essentially
depends on the parameter (\ref{Qdef}) which in the absence of the Lorentz
Chern-Simons coupling ($\alpha_{\rm L} = 0$) reduces to $Q = \Lambda^{\rm eff}
+ \ell\varepsilon_0/\chi$ (note that $\beta = - 1/\chi$ then). The explicit 
solution of (\ref{eq0}) and (\ref{dr}) can be conveniently written in the form
\begin{equation}
R = R_1R_2,\qquad 2r = {\frac {c_0}{\sqrt{c_0^2 + \ell\varepsilon_2/\chi}}}
\,\ln{\frac {R_1}{R_2}},
\end{equation}
where the functions $R_1(t)$ and $R_2(t)$ are determined by:
\begin{equation}
R_1 = \left\{\begin{array}{ccl}
{\frac 1 {\sqrt{Q}}}\,\sinh(\sqrt{Q}\,t), & \ {\rm if}\ & Q > 0,\\ 
{\frac 1 {\sqrt{|Q|}}}\,\sin(\sqrt{|Q|}\,t), & \ {\rm if}\ &  Q < 0,\\ 
t, & \ {\rm if}\ &  Q = 0.\end{array}\right.
\end{equation}
\begin{equation}
R_2 = \left\{\begin{array}{ccl}
{\frac {\ell\varepsilon_1} {\chi\sqrt{Q}}}\,\sinh(\sqrt{Q}\,t) + 
2\sqrt{c_0^2 + \ell\varepsilon_2/\chi}\,\cosh(\sqrt{Q}\,t), & \ {\rm if}
\ & Q > 0,\\ {\frac {\ell\varepsilon_1} {\chi\sqrt{|Q|}}}\,\sin(\sqrt{
|Q|}\,t) + 2\sqrt{c_0^2 + \ell\varepsilon_2/\chi}\,\cos(\sqrt{|Q|}\,t), 
& \ {\rm if}\ &  Q < 0,\\ {\frac {\ell\varepsilon_1}\chi}\,t + 2\sqrt{c_0^2
+ \ell\varepsilon_2/\chi}, & \ {\rm if}\ &  Q = 0.\end{array}\right.
\end{equation}

As a result, the metric takes the form of the Heckmann-Sch\"ucking type 
anisotropic cosmological solution: 
\begin{equation}
ds^2 = -\,dt^2 + \left(R_1^{p_1}R_2^{1-p_1}\right)^2dx^2 + 
\left(R_1^{p_2}R_2^{1-p_2}\right)^2dy^2.
\end{equation}
Here we introduce the two constant parameters which satisfy the Kasner type 
conditions:
\begin{equation}
p_1 + p_2 = 1,\qquad p_1^2 + p_2^2 = 1 - \xi/2.
\end{equation}
Like in four dimensions \cite{kam}, the ``non-Kasner" parameter 
\begin{equation}
\xi := {\frac {\ell\varepsilon_2}{\chi c_0^2 + \ell\varepsilon_2}}
\end{equation}
vanishes when the stiff matter component is absent. 

The corresponding Riemannian curvature 2-form reads:
\begin{eqnarray}
\widetilde{R}^{\widehat{0}\widehat{1}} &=& \left(- Q + {\frac {4p_1p_2
(c_0^2 + \ell\varepsilon_2/\chi)}{R^2}}\right)\vartheta^{\widehat{0}}\wedge
\vartheta^{\widehat{1}},\\ \widetilde{R}^{\widehat{0}\widehat{2}} &=& 
\left(- Q + {\frac {4p_1p_2(c_0^2 + \ell\varepsilon_2/\chi)}{R^2}}\right)
\vartheta^{\widehat{0}}\wedge\vartheta^{\widehat{2}},\\ \widetilde{R}^{
\widehat{1}\widehat{2}} &=& \left(- Q - {\frac {4p_1p_2 (c_0^2 + \ell
\varepsilon_2/\chi)}{R^2}} - {\frac {\ell\varepsilon_1/\chi} R}\right)
\vartheta^{\widehat{1}}\wedge\vartheta^{\widehat{2}}.
\end{eqnarray}

In the absence of the dust and the stiff matter contributions (i.e., when
both $\varepsilon_1 = 0$ and $\varepsilon_2 = 0$), we recover the anisotropic 
de Sitter geometry studied in the previous subsection.

\section{Conclusions}\label{DC}

In this paper, we have derived the effective 3D Einstein theory which is
equivalent to the Mielke-Baekler model for all matter sources with the
vanishing dynamical spin current. The Lagrangian of the model includes
the standard Hilbert-Einstein term plus the translational and the rotational 
Chern-Simons terms. It is shown that the purely Chern-Simons gravitational
theory (when the Hilbert-Einstein term is absent) are only consistent in
vacuum, whereas the nontrivial energy-momentum current leads, in general, 
to the inconsistencies.

The general formalism is applied to the cases of electromagnetic and the
perfect fluid sources. The new exact solutions of the 3D gravity theory
are obtained: the gravitational $pp$-waves for the Maxwell source, and
the black hole type solutions and the anisotropic cosmological solutions
for the perfect fluid matter.

\bigskip
{\bf Acknowledgments}. This work was supported by the Deutsche 
Forschungsgemeinschaft (Bonn) with the grant HE~528/20-1. I thank Friedrich
Hehl for reading the paper and discussions.

\end{document}